\documentclass[12pt, a4paper]{article}

\begin{document}
\title{Parrondo Games and Quantum Algorithms}
\author{Chiu Fan Lee\thanks{c.lee1@physics.ox.ac.uk} 
\ and \ Neil Johnson\thanks{n.johnson@physics.ox.ac.uk}
\\
\\ Center for Quantum Computation and Physics Department \\ Clarendon
Laboratory, Oxford University \\ Parks Road, Oxford OX1 3PU, U.K.}

\maketitle

\abstract{We pursue the possible connections between classical games and
quantum computation. The Parrondo game is one in which a random combination
of two losing games produces a winning game. We introduce novel
realizations of this Parrondo effect in which the
player can `win' via random reflections and rotations of the state-vector,
and connect these to known quantum algorithms.}

\newpage
The possibility of `winning by losing' is obviously appealing. This
was the idea behind the Brownian ratchet studied by Feynmann \cite{Feyn}.
Although one cannot `win' against the Second Law of Thermodynamics, a
discrete-time version of this phenomenon can be interpreted as a
gambling game. In particular, Parrondo and co-workers showed that two losing
classical games, A and B, could be combined to create a winning game
\cite{Parr}.   In a separate development, the field of
quantum games is currently attracting much attention because of the
underlying theme of information processing \cite{qgame}. 
But of what possible use are such quantum games? It is
hoped they might offer some deeper insight into quantum complexity and,
specifically, the design of new quantum algorithms \cite{Meyer,Nature}. 
However this hope is still little more than speculation.

In this paper, we consider the Parrondo
effect in the  quantum domain and provide an interesting connection between
this phenomenon  and known quantum algorithms.
The Parrondo effect allows a player to profit from randomness (i.e.
noise) - hence it has possible practical relevance in the stochastic control
of quantum algorithms
\cite{port} and in controlling quantum decoherence
\cite{stoch}.  Two very different implementations of the Parrondo idea
have recently been proposed for quantum games, with spatial state-dependence
\cite{Meyer2} and historical
state-dependence \cite{Flitney}.
We note however that the latter implementation did not consider random
switching between the two losing quantum games; moreover, neither
implementation made an explicit connection to quantum algorithms. In this
paper we make such a connection, by explicitly focussing on the effect of
random rotations and reflections on the relevant state-vector.

We start by reviewing the Parrondo effect, but using a novel realization
in terms of rotating vectors. Such a viewpoint makes sense given that our
objective is to connect up with quantum algorithms: such
quantum algorithms can be thought of geometrically in terms of
rotation/reflection operations on the state-vector within the Bloch sphere
\cite{Chuang}. Comparing to
previous realizations of classical Parrondo's games, our
rotating-vector realization produces a much higher winning rate for the
combined game
$(A\oplus B)$ even though the losing rates for games A and B are greater.
This improvement is achieved at the  expense of having to keep track of more
states. We note that Parrondo games are one-person games, hence the
present discussions only refer to `the player' (in contrast to the quantum
games of Refs.
\cite{qgame,Meyer}). 

\noindent {\em Game A:\ } Consider a wheel with a vector 
drawn from the center to the circumference, i.e. the vector is a radial
line.  Suppose the vector is originally vertical (i.e. $\theta=0$) and the
player plays by calling a robot (A) to rotate the  wheel. The robot can only
rotate the wheel by
$0$, $2\pi/3$ or
$4\pi/3$ radians, with equal probabilities. The player wins if the vector
ends up in the  upper-half of the circle (i.e. $-\pi/2\leq \theta\leq
\pi/2$) and he loses otherwise. The game is continued by rotating the wheel
from the previous position, i.e. without restoring the vector to the
vertical position. The stationary states are such that the vector will
end up at $\theta=0$, $2\pi/3$ or
$4\pi/3$
with equal probabilities. Therefore
this game is losing for the player and the rate of losing is $1/3$. In
Parrondo's original game, the losing rate is smaller (i.e. $-2\epsilon$ where
$\epsilon\ll 1$).

\noindent {\em Game B:\ } This is the same as game
A, except that the robot (B) can now only rotate the wheel by 
$0$, $2\pi/7$,
$4\pi/7$, $6\pi/7$, $8\pi/7$, $10\pi/7$, $12\pi/7$,  with
equal probabilities. Similar analysis as that for game A shows the
player's losing rate is $1/7$. In
Parrondo's original game, the losing rate is again smaller 
(i.e. $-11\epsilon/5$ where 
$\epsilon\ll 1$).

\noindent {\em Game $A\oplus B$:\ } The player now plays a combined game
in which he randomly selects either A or B at each timestep. 
Operationally, one of the robots A or B is
selected at random to rotate the wheel at each timestep. Simple geometric
analysis shows that the vector can now end up in
$3\times 7=21$ different orientations, $11$ of which are
winning. 
The corresponding  $21\times21$ transition matrix is doubly-stochastic and
so the stationary distribution will be equally distributed among these $21$
positions. Therefore the player now wins with probability
$11/21\approx 0.5238 >1/2$. In
Parrondo's original game, the winning rate was $1/80 -
21\epsilon/10$ as compared to the present, larger rate of $1/21$. 
It turns out there is nothing special
about the numbers $3$ and $7$ chosen for this implementation. The games A
and B are originally losing simply because $3=7=3 \bmod 4$, and the combined
game becomes winning because
$3\times 7 =1 \bmod 4$. Therefore, the above vector-rotating implementation
of Parrondo's effect works equally well for all $m,n$ such that $(m,n)=1$ and
$m=n=3\bmod 4$. By the same method, we can therefore construct two losing
games with rates
$-1/m<0$ and
$-1/n<0$ such that when they are combined at random, we obtain a
winning game with rate
$1/mn>0$. One could also extend the Parrondo scheme to include random
combinations of  {\em any even} number of games. We therefore conclude that 
random rotations of a vector can be used to `win', in the same spirit as the
original Parrondo effect. 

Can we construct similar situations in the quantum domain in which
we can also `win' using randomly-chosen symmetry operations on the
state-vector, and can we expect these `games' to have a close connection with
quantum algorithms? After all, a quantum state is  by definition a
normalized vector in the corresponding  Hilbert space and a quantum
computation can be seen as a series of rotations/reflections applied to the
original state. We now proceed to show that 
this is indeed the case, and that there is indeed an
interesting connection to quantum algorithms. We speculate that such a
connection could, in principle, provide a starting point for
uncovering new quantum algorithms - the systematic generation of
such algorithms is one of the great outstanding problems in quantum
computing.

Our first example of the Parrondo effect in the quantum domain, is related
to the Bernstein-Vazirani algorithm for guessing a number
\cite{Bernstein,Meyer}. Any rotation can be seen as a reflection about a
specific plane, hence for convenience our discussions will be couched in the
language of reflections. The goal of the game is to obtain
(i.e. measure with a high probability) a final state
$|\alpha\rangle$ starting from an initial state $|0\cdots 0\rangle$
\cite{Meyer}. The player must play this
game for as few timesteps as possible: he has at his disposal
only two types of operation, each of which corresponds to an oracle.
The first corresponds to the following
operation:
\[ \hat{O}|x\rangle = (-1)^{x\cdot\alpha}|x\rangle
\]
where $x\cdot\alpha$ denotes the bit-wise inner product modulo 2
\cite{Chuang}. However there is a catch for the player: this oracle is
`noisy'. In particular,  the operation
$\hat{O}$ is only performed for {\em half} the computational basis, hence
\[ \hat{O}|x\rangle = \left\{ \begin{array}{l}  (-1)^{x\cdot\alpha}|x\rangle\\
|x\rangle
\end{array} \right.
\] with equal probability. 
The second oracle is reliable, and corresponds to applying
the operator $H^{\otimes n}$. 
Following Ref. \cite{Bernstein}, if both oracles were noiseless then the
player could obtain
$|\alpha\rangle $ in only three timesteps via the sequence $H^{\otimes n} 
\hat{O}
H^{\otimes n}$.   Suppose instead the player plays this sequence using
the noisy oracle. Applying the noisy
$\hat{O}$ operator is the same as  reflecting $|\psi \rangle 
:=\sum^{2^n-1}_{x=0}
\frac{1} {\sqrt{2^n}}|x\rangle$ about all $|y\rangle$ such that $y\cdot
\alpha =1$ at random. If the player were to play this `game' of reflection
for each
$y$ separately, it would only increase his chances of measuring $\alpha$ by
a very small amount.  Indeed, the chance of measuring $\alpha$ becomes
${\sc O}(1/2^n)$. Now we explore what would happen if the player combines all these
`games' together. Assuming
$\alpha
\neq 0$, then $\#\{y | y\cdot \alpha =1 \}= 2^{n-1}$ hence
\[ \hat{O}H^{\otimes n}|0\cdots 0\rangle
=\frac{1}{\sqrt{2^n+2^n}}[\sum^{2^n-1}_{x=0} (-1)^{x\cdot
\alpha}|x\rangle+2|y_1\rangle + \cdots + 2|y_{2^{n-2}}\rangle].
\]
It can be shown that the probability of measuring $\alpha$ is now given by
\[ |\langle \alpha| H^{\otimes n}
\hat{O}H^{\otimes n} |0 \cdots 0\rangle|^2 >1/8
\] which represents a very substantial improvement over the previous value
of ${\sc O}(1/2^n)$.
In the spirit of the Parrondo effect, well-controlled 
randomness has been exploited for profit such that the player now `wins'.

We now present a quantum
Parrondo game which can be viewed as a
vector-rotating game, and is related to Grover's search algorithm
\cite{Grover}. The player's goal is to obtain (i.e. measure with a high
probability) a fixed, unknown number
$\alpha$ in as few timesteps as possible, where $0
\leq
\alpha \leq 2^n-1$. The initial
state has the form
$|\psi\rangle=\sum^{2^n-1}_{x=0}
\frac{1} {\sqrt{2^n}}|x\rangle$.  In this game, an infinite sequence of
operators
$
\hat{O}_1
\cdots \hat{O}_m \cdots$ will  be applied to $|\psi \rangle$. The player
decides when to stop the sequence, i.e. he has the freedom
to choose
$m$ such that $|\psi_f\rangle = \hat{O}_m \cdots \hat{O}_1|\psi\rangle$. The
payoff is then determined by a measurement in the computational basis of
$|\psi_f\rangle$. The game is winning if the player possesses a strategy
that wins with probability $>1/2$,  and is losing
otherwise. This game incorporates strategic  moves, since the
set of strategies used by the player to decide the duration of the game 
are equivalent to the set of natural numbers
${\bf N}$.

\noindent {\em Game A:\ } Here $\hat{O_i} = \hat{A}$ for all $i$, 
 where $\hat{A}(|x \rangle ) = (-1)^{\delta_{x \alpha}} |x \rangle$. 
Geometrically, $\hat{A}$ reflects the vector $|\psi\rangle$ about $|\alpha
\rangle$. Since $\hat{A}^2=I$, the player's freedom in choosing when to
stop the game will always reduce to just one of the following two
scenarios: 
$|\psi_f \rangle = \hat{A} |\psi \rangle$ or
$|\psi_f \rangle = |\psi \rangle$. Unfortunately for the player,
the payoff $|\langle \alpha | \hat{A}|\psi \rangle|^2 = |\langle \alpha |\psi
\rangle|^2 = \frac{1}{2^n}$ which is less than $1/2$ for $n
\geq2$. Therefore the player does not possess a winning strategy,
hence game A is losing for him. 

\noindent {\em Game B:\ }  Here $\hat{O_i}=\hat{B}$ for all $i$, where
$\hat{B}:=2|\psi \rangle \langle \psi | - I$. 
Geometrically, $\hat{B}$ reflects $|\psi \rangle$ about itself.
Again, the player has the freedom to decide how many $\hat{B}$ are applied
to the input state before measurement. However since $\hat{B}|\psi \rangle
=|\psi
\rangle$, the player can have no influence in determining the payoff
in this game. The  game is hence losing for him because the payoff $|\langle
\alpha |
\psi
\rangle|^2 = \frac{1}{2^n}$ which is less than $1/2$.

\noindent {\em Game $A\oplus B$:\ } The player combines games A and B at
random. By this we mean $\hat{O}_i = \hat{A}$ or $\hat{B}$ with equal
probability, and the player is told the sequence as it unfolds. Once
again, the player has the freedom to decide when to stop the sequence and
hence do the measurement.
Since $\hat{A}^2=\hat{B}^2 =I$ and
$\hat{B}|\psi \rangle=|\psi\rangle$, any given finite
sequence
$\hat{O}_i$ will always produce a final state with the following form: 
$|\psi_f\rangle =(\hat{B})\hat{A}\hat{B}\cdots \hat{A}\hat{B}\hat{A}|\psi
\rangle$. For all $k\in {\bf Z}$,  there will {\em almost
surely} \cite{event} be an $m(k)\in {\bf Z}$ such that
\begin{eqnarray*} |\psi_f\rangle &=& \hat{O}_{m(k)} \cdots
\hat{O}_1|\psi\rangle \\ &=&
\overbrace{(\hat{B}\hat{A})\cdots (\hat{B}\hat{A})}^{k}|\psi\rangle.
\end{eqnarray*} 
A winning strategy for the player is therefore to choose $m(k)$ such that 
$k = \lceil \pi \sqrt{2^n}/
 4\rceil$. 
By stopping after the $m(k)$-th operation, it can be shown that the player
will then win with  probability $>1/2$. As such an $m(k)$ exists almost
surely \cite{event}, we see that this combined game is winning for the
player.  It can also be seen that $\hat{B}
\circ
\hat{A} =
\hat{G}
$ where $\hat{G}$ is Grover's operator \cite{Grover,Chuang}. 
Thus we have produced a Parrondo effect whereby the player benefits from
randomness to win, and have connected this effect to a well-known quantum
algorithm. 

Throughout this paper we have discussed combinations of games, each
of which constitutes particular strategies. One could argue that the {\em
combined} game ($A\oplus B$) is the true game, and that the `games'
A and B are just strategies in this larger game chosen in some
(random) sequence. This is just a semantic point, and could also
be made against much of the Parrondo game literature. In order to make direct
connection with this literature and community, we prefer to regard the
combined game
$A\oplus B$ as a (random) sequence of games A and B rather than a (random)
sequence of strategies. The physical results and implications are unchanged
by such shifts in verbal definitions.

To summarize, we have discussed connections between quantum games and quantum
algorithms whereby the performance can be improved by profiting from
randomness. The counter-intuitive conclusion is that such randomness/noise
can be of direct use in the quantum regime. We hope that the present results
serve to stimulate further work in this on-going field. 

\vskip0.5in CFL thanks NSERC (Canada), ORS (U.K.) and
Clarendon Fund (Oxford) for financial supports. NFJ thanks EPSRC
for a travel grant, and is grateful to L. Quiroga, F. Rodriguez and D.
Abbott for discussions.


\begin{thebibliography}{99}
\bibitem{event} An event $E$ occurs almost surely if $\forall \epsilon >0,
{\rm Pr}(E) >1-\epsilon$.
\end{thebibliography}
\end{document}